\documentclass[10pt,a4paper]{article}

\usepackage[utf8]{inputenc}
\usepackage{amsmath,amsfonts,amssymb,amsthm, multirow}
\usepackage{graphicx}
\usepackage{fullpage}
\usepackage{hyperref}
\usepackage{natbib}
\usepackage{color}
\usepackage{tabularx}
\usepackage{amsfonts,amssymb,amsmath,amscd,latexsym,dsfont}
\usepackage{enumitem}

\usepackage{graphicx}
\usepackage{subfigure}
\usepackage{natbib}

\newcommand{\bomega}{\boldsymbol{\omega}}
\newcommand{\boeta}{\boldsymbol{\eta}}

\newcommand{\bth}{\boldsymbol{\theta}}
\newcommand{\balpha}{\boldsymbol{\alpha}}

\newcommand{\bpi}{\boldsymbol{\pi}}
 \newcommand{\bm}{\boldsymbol{m}}

 \newcommand{\MM}{\mathcal{M}}
\newcommand{\bG}{\boldsymbol{G}} 

\newcommand{\bx}{\boldsymbol{x}} 
\newcommand{\tx}{\textbf{x}}
\newcommand{\tz}{\textbf{z}}
 \newcommand{\bz}{\boldsymbol{z}}
\newcommand{\ttt}{\textbf{t}} 

\newcommand{\pdf}{p}
{\it}{\rm}
\newtheorem{ass}{Assumption}
\newtheorem{theorem}{Theorem}

\newcommand{\argmax}{\mathop{\mathrm{arg\,max}}}

\title{Non-parametric Multi-Partitions Clustering}

\author{Marie du Roy de Chaumaray and Vincent Vandewalle}

\begin{document}

\maketitle

\begin{abstract}
In the framework of model-based clustering, a model, called multi-partitions clustering, allowing several latent class variables has been proposed. 
This model assumes that the distribution of the observed data can be factorized into several independent blocks of variables, each block  following its own mixture model.
In this paper, we assume that each block follows a non parametric latent class model, {\it i.e.} independence of the variables in each component of the mixture with no parametric assumption on their class conditional distribution. The purpose is to deduce, from the observation of a sample, the number of blocks, the partition of the variables into the blocks and the number of components in each block, which characterise the proposed model. By following recent literature on model and variable selection in non-parametric mixture models, we propose to discretize the data into bins. This permits to apply the classical multi-partition clustering procedure for parametric multinomials, which are based on a penalized likelihood method (\emph{e.g.} BIC). The consistency of the procedure is obtained and an efficient optimization is proposed. The performances of the model are investigated on simulated data.



\end{abstract}


\maketitle

\section{Introduction}


Finite mixture models allows to perform clustering by modelling the distribution of the variable and identifying each component of the mixture as a cluster \citep{McL00,mcnicholas2016mixture,bouveyron2019model}. This has permitted to perform the clustering of a wide range of data (continuous, categorical, functional, mixed) by adapting the class conditional model to each kind of data. Most of the development of mixture models have been performed in a parametric framework where it is possible to perform consistent model selection such has choosing the number of cluster \citep{keribin2000consistent}. However assuming a parametric model can be too restrictive to encompass the variety of cluster shapes. Thus non-parametric models have been developed (see \citet{chauveau2015semi} for a review), they typically only assume class conditional independence of the variables of the cluster making no parametric assumption on the class conditional univariate distribution. Then they are able to perform estimation with an EM-like algorithm \citep{BenagliaJCGS2009} or by maximizing the smoothed log-likelihood \citep{LevineBiometrika2011}. In this setting the choice of the number of cluster is a difficult issue for which \citet{MMfull} have recently proposed a consistent solution based on the discretization of the variables.

One limitation of finite mixture models is that they try to summarize the heterogeneity of the data by only one categorical variable. However, in the area of massive data, with individuals described by possibly thousands of variables the whole heterogeneity in the data cannot be described by only one latent variable. 
It can for instance be the case if variables related to some focus are more present than variables related to another one. Thus model-based clustering approaches has been developed to possibly handle several latent class variables, in what we latter call multiple partitions clustering (see \citet{rodriguez2022multipartition} for a recent review on the subject). Assuming several latent variables in the model can be performed in three principal ways. It can be performed by assuming that the distribution factorises in independent blocks of variables, the heterogeneity in each block being explained by a latent clustering variable \citep{GALIMBERTI2007520, MMVV}. It can be performed by assuming several classifying linear projections of the variables, each one being explained by a latent cluster variable \citep{attias1999independent, vandewalle2020multi}. Finally, several latent variables can be considered in more complex dependence structure such as trees \citep{poon2013model}, Bayesian networks \citep{rodriguez2022multipartition}, or multilayer (potentially deep) discrete
latent structure \citep{gu2021bayesian}. Multiple partition clustering induces an additional complexity compared to standard clustering, since a lot of structure parameters need to be learned (number of blocks, repartition of the variables in blocks, number of modalities of each latent variables, structure of the network). Since the above multiple partitions models fall in the parametric framework all these parameters can be selected using penalized (such as BIC) or integrated likelihood based criteria (such as MICL). The difficulty of this search depends on the complexity of models which is assumed. For instance in the continuous setting \citet{Galimberti2017} propose a extension of \citet{GALIMBERTI2007520} where many possible roles of the variables need to be considered, thus needing a lot of computation even for the re-affectation of only one variable. Contrarily \citet{MMVV} propose a very simple model in which variables are grouped into independent blocks and each block of variables is a assumed to follow a mixture model with the class conditional independence assumption in a parametric setting, they were able to propose a modified EM algorithm allowing to update independently the affectation of the variables to the block at each step of the algorithm. Let notice that this model is an extension of the approaches proposed by \citet{marbac2017variable,marbac2017variable2,marbac2020variable} in the framework of variable selection in clustering, where only two blocks are considered, {\emph i.e.} one block of classifying variables assuming conditional independence, and one block of non classifying variables assuming total independence.

In this paper we propose a non parametric extension of the approach proposed by \citet{MMVV}. In order to solve the difficulty of the model choice, we first discretize the data, where the granularity of the discretization depends on the number of data as in \citet{MMfull}. In this case the non parametric multipartition mixture model becomes a parametric one where each block follows a mixture of product of multinomial distributions. This model is a particular case of the model proposed by \citet{MMVV}, thus we are able to perform an efficient model search and parameters estimation by optimizing a penalized likelihood. Moreover following the same lines as \citet{MMfull} we are able to prove the consistency of this procedure. Once the issue of model selection is solved, we propose to refit a non parametric latent class model on each block of variables with fixed number of clusters in order to limit the loss of information caused by the discretization of variables and thus to obtain a more accurate estimation of the partitions. 

The outline of the paper is the following: Section \ref{sec:model} introduces the non-parametric multiple partitions model, Section \ref{sec:bin} explains how the data are discretized, while Section \ref{sec:estim} presents the estimator of the model along with its consistency and with the modified EM algorithm used for model selection.
Finally, Section \ref{sec:experiments} illustrates the good performances of our procedure on simulated data.

\section{Multiple partitions mixture model} \label{sec:model}
\subsection{The underlying model}

Data to cluster $\tx=(\bx_1,\ldots,\bx_n)$ are composed of $n$ observations $\bx_i=(x_{i1},\ldots,x_{id})$ described by $d$ variables potentially of different types (\emph{i.e.,} each variable can be continuous or categorical).
Observations are assumed to independently arise from a multiple partitions model (MPM) which considers that variables are grouped into $B$ independent blocks. 
The blocks of variables are defined by $\bomega=(\omega_{1},\ldots,\omega_d)$, where $\omega_{j}=b$ indicates that variable $j$ belongs to block $b$. The set of the indexes of variables which belong to the same block $b$ is denoted by $\Omega_b=\{j: \omega_{j}=b\}$. 
Moreover, MPM considers that the variables of block $b$ follow a $G_b$-components mixture assuming within-component independence. This assumption permits to write the conditional density of those variables as the product of $|\Omega_b|=$card$(\Omega_b)$ univariate densities. A model $\bm$ is thus given by the number of blocks $B$, the repartition of the variables in each block $\bomega$ and the numbers of components $\bG=(G_1,\ldots,G_B)$ in the mixture-model driving each block.  Let $\bx_{i\{b\}}=(x_{ij};j\in\Omega_b)$ be the vector of observed variables of block $b$. The probability distribution function (pdf) of $\bx_i$, for a model  $\bm=(B, \bG, \bomega)$, is thus given by
\begin{equation}
  \pdf (\bx_i | \bm, \bth ) = \prod_{b=1}^B   \pdf_b (\bx_{i\{b\}} | \bm, \bth )  
  \text{ with }
   \pdf_b (\bx_{i\{b\}} | \bm, \bth ) = \sum_{g = 1}^{G_b} \pi_{bg} \prod_{j \in \Omega_b} \eta_{gj}(x_{ij}), \label{eq:model}
 \end{equation}
where  $\bth= (\bpi, \boeta)$ groups the model parameters, with $\bpi=(\bpi_{b};b=1,\ldots,B)$ being the proportions of the components  in each mixture, with $\bpi_b=(\pi_{b1},\ldots,\pi_{bG_b})$, $\pi_{bg}>0$ and $\sum_{g=1}^{G_b} \pi_{bg}=1$, and $\boeta$ grouping the  univariate densities $\eta_{gj}$, which are infinite dimensional parameters for the continuous variables. We denote by $\Theta_{\bm}$ the set of all possible parameters $\bth$ associated with a given model $\bm$. And we denote by $\pdf_0$ the pdf under the true model $\bm_0$ and the true parameter $\bth_0$, \emph{i.e.} $\pdf_0 = \pdf (.| \bm_0, \bth_0)$.

\subsection{Resulting partition of the data}
MPM provides $B$ partitions among the observations (one partition per block of variables). 
The partition of block $b$ is denoted by  $\tz_{b} = (\bz_{1b}, \ldots, \bz_{nb})\in \mathcal{Z}_{G_b}$, where $\mathcal{Z}_{G_b}$ is the set of all partitions of $n$ elements into $G_b$ clusters, and $\bz_{ib} = (z_{ib1},\ldots,z_{ibG_b})$ with $z_{ibg} = 1$ if observation $i$ belongs to cluster $g$ for block $b$ and $z_{ibg} = 0$ otherwise. The multiple partitions $\tz=(\tz_1,\ldots,\tz_B)$ for model $\bm$ thus belongs to  $\boldsymbol{\mathcal{Z}}_{\bm} = {\bf \mathcal{Z}}_{G_1}\times\ldots\times\mathcal{Z}_{G_B}$. 
 

\subsection{Model and parameters identifiability} \label{sec:identif}
In this part, we explain why model \eqref{eq:model} is identifiable up to a switching of the component labels and a change in the order of the blocks. We need the following Assumptions.

\begin{ass} \label{ass:identifiability}
\begin{enumerate}[label=(\roman*)]
    \item \label{ass:id1} The number of variables is at least three in each block and all the proportions in the mixtures are not zero, \emph{i.e.} for any $b=1, \ldots B$, $|\Omega_b| \geq 3$ and, for any $g= 1, \ldots, G_b$,  $\pi_{bg}>0$. 
    \item \label{ass:id2} For each $b=1, \ldots, B$, there exists $\Upsilon_b \subseteq \Omega_b$ such that $|\Upsilon_b|=3$ and for any $j\in\Upsilon_b$ the  univariate densities $\{\eta_{gj}; g=1, \ldots, G_b\}$ are linearly independent.
    \item \label{ass:id3} For each $b=1, \ldots, B$, $\pdf_b$ cannot be decomposed in a product of densities.
\end{enumerate}
\end{ass}

Suppose $B$ and $\bomega$ fixed and known, which means that the block structure of the variables is given, then, under Assumptions \ref{ass:identifiability}\ref{ass:id1} and \ref{ass:id2}, the model parameters $\bth$ and the numbers of components $\bG$ are identifiable (see \cite{All09}). 

We define $B$ as the greatest integer which permits to decompose $\pdf(\cdot|\bm, \bth)$ as the product $\prod_{b=1}^B   \pdf_b (\cdot | \bm, \bth )$. Assumption \ref{ass:identifiability}\ref{ass:id3} thus ensures the identifiability of $B$ and of the associated partition of the variables $\bomega$. 

Note that Assumption \ref{ass:identifiability}\ref{ass:id3} means that the variables belonging to the same block are not independent.



\color{black}


\section{Model selection via bin estimation}\label{sec:bin}

\subsection{The discretized model used for estimation}
If all the variables are assumed to be continuous, the method is based on the discretization of each variable $j$ into $R$ non-overlapping  bins $I_{Rj1},\ldots,I_{RjR}$ such that $\cup_{r=1}^R I_{Rjr}=\mathcal{X}_j$ and for any $(r,r')$ with $r\neq r'$, $I_{Rjr}\cap I_{Rjr'}=\emptyset$. We denote by  $\sigma_{Rjr}$ with $r\in\{1,\ldots,R\}$, the indicator functions of each bin, defined by $\sigma_{Rjr}(x_{ij})=1$ if $x_{ij}\in I_{Rjr}$ and $\sigma_{Rjr}(x_{ij})=0$ if $x_{ij}\notin I_{Rjr}$, and we denote by $|I_{Rjr}|$ the Lebesgue measure of the bin $I_{Rjr}$. Alternatively, one could use different numbers of bins per variables, which is not done in the following for ease of reading. The number of bins as well as their location is known and fixed upstream. Conditions to ensure the good property of the procedure and ways to choose those items will be given in the next section. 

The discretized variables of block $b$ follow a latent class model where each component is a product of $|\Omega_b|$ multinomial distributions each having $R$ levels. Therefore, the discretized pdf of the subject $i$ for the variables in block $b$ is given by
\begin{equation}
\pdf_{Rb} (\bx_{i\{b\}} | \bm, \bth_{Rb} ) = \sum_{g = 1}^{G_b} \pi_{bg} \prod_{j \in \Omega_b} 
\prod_{r=1}^R \left(\frac{\alpha_{Rgjr}}{|I_{Rjr}|}\right)^{\sigma_{Rjr}(x_{ij})},\label{eq:modeldiscret}
\end{equation}
where $\bth_{Rb}=(\bpi_b, \balpha_{Rb})$ groups the component proportions $\pi_{bg}$ and the probabilities $\alpha_{Rgjr}$  that one subject arisen from component $g$ takes level $r$ for the variable $j$ when this variable is discretized into $R$ bins. The division by $|I_{Rjr}|$ stands for the histogram approximation of the class conditional univariate densities. The parameter space is given by the product of simplexes $S_{G_b} \times S_R^{G_b|\Omega_b|}$. Note that $\pdf_{Rb}$ is an approximation of $\pdf_b$ and that this approximation becomes more accurate as $R$ tends to infinity.  We deduce a discretized version $\pdf_R$ of the pdf $\pdf$ of subject $i$, which is given by
\begin{equation*}
\pdf_R(\bx_i | \bm, \bth_{R} ) = \prod_{b=1}^B   \pdf_{Rb} (\bx_{i\{b\}} | \bm, \bth_{Rb} ),
\end{equation*}
where $\pdf_{Rb}$ is given by Equation \eqref{eq:modeldiscret} and $\bth_R$ groups all parameters of each block $\bth_{R1}, \ldots, \bth_{RB}$.
The discretized version  of the true density $\pdf(.|\bm_0, \bth_0)$ will be denoted by $\pdf_{0,R}$.
Note that, by construction, the discretization induces a loss of identifiability concerning the parameters $\alpha_{Rgjr}$.

\subsection{Dealing with mixed-type data}
The model permits to deal with mixed-type data also, where some of the variables are categorical. Indeed, we only discretize the variables which are continuous.  

\section{Model selection via penalized likelihood}\label{sec:estim}

The set of competing models $\MM$ is given by 
\begin{equation}
\MM = \{ \bm=(B,\bG,\bomega);  B  \leq B_{\max}, \; \forall b, j \; G_b \leq G_{\max},  \omega_j \in \{1,\ldots,B\}, |\Omega_b| \geq 3  \}, \label{eq:competing}
\end{equation}
where $B_{\max}$ is the maximum number of blocks and $G_{\max}$ is the maximum number of components within a block.
 For a given model $\bm$, the parameters $\bth_R$, constituted by all the $\pi_{bg}$ and all the $\alpha_{Rgjr}$, are unknown and must be estimated from the observations. The parameter space for a given model $\bm$ is denoted by $\Theta_{R,\bm}$ and is the following product space 
\begin{equation}\label{eq:paramspace}
\Theta_{R, \bm} = \prod_{b=1}^B S_{G_b} \times S_R^{G_b |\Omega_b|},
\end{equation} 
where $S_K = \{\boldsymbol{u}\in[0,1]^K : \;\sum_{k=1}^K u_k=1\}$  designates the simplex of size $K$.
We decide to choose the model which maximizes the penalized log-likelihood of the discretized version, for some well-chosen penalty $a_{n, \bm, R}$. For instance, it corresponds to the BIC \citep{Schwarz:78} if $a_{n, \bm, R}$ is equal to $\nu_{\bm} \ln(n)/2$ where $\nu_{\bm}$ is the complexity of model $\bm$. According to Equation \eqref{eq:paramspace}, $$\nu_{\bm}= \sum_{b=1}^B (G_b-1) + (R-1) G_b |\Omega_b|.$$

\subsection{Model inference}

For a sample $\tx$ and a model $\bm$, the observed-data log-likelihood is defined by
\begin{equation}\label{eq:loglik}
\ell(\bth_R|\bm,\tx)=\sum_{i=1}^n \sum_{b=1}^B  \ln \pdf_{Rb} (\bx_{i\{b\}} | \bm, \bth_{Rb} ),
\end{equation}
and we obtain its penalized version $\ell_{\text{pen}}$ by subtracting the penalty term
\begin{equation}
\ell_{\text{pen}}(\bth_R|\bm,\tx) = \ell(\bth_R|\bm,\tx) - a_{n, \bm, R}
\end{equation}
Model selection with penalized likelihood consists in maximizing $\ell_{\text{pen}}$ with respect to $\bth_R$ and then with respect to $\bm$. Thus, the selected model is given by
\begin{equation}\label{eq:mhat}
\widehat{\bm}_{n, R} = \argmax_{\bm \in \MM} \max_{\bth_R \in \Theta_{R,\bm}} \ell_{pen}(\bth_R|\bm,\tx).
\end{equation}
Note that, thanks to the conditional independence, we can sum over $i$ before summing over $b$ in equation \eqref{eq:loglik}, which means that the maximization can be done separately in each block.

 In practice, to avoid numerical issues, we introduce a threshold $\varepsilon$ such that the parameter space becomes $\Theta_{R,\bm ,\varepsilon}= \prod_{b=1}^B S_{G_b,\varepsilon} \times S_{R,\varepsilon}^{G_b|\Omega_b|}$, with $\varepsilon>0$ being the minimal value of all the elements defined in the simplexes, $\textit{i.e.} , \: S_{K,\varepsilon}=\{\boldsymbol{u}\in ] \varepsilon, 1]^K: \; \; \sum_{k=1}^K u_k=1 \}$.   Note that the use of such a threshold is quite usual in this framework (see \citet{Toussile2009, Bontemps2013, MMfull}). 
 Under the condition that $R \varepsilon$ tends to zero as $R$ goes to infinity and $\varepsilon$ to zero, the parameter space $\Theta_{R,\bm, \varepsilon}$ converges to the whole parameter space  
 $\Theta_{R,\bm}$ as $\varepsilon$ tends to zero.
 According to  the assumptions on the growth rate of $B$ which will be stated by Assumption~\ref{ass:intervals}\ref{ass:vitesseB} in the next section, it is sufficient to set $\varepsilon^{-1}=O(n^{\alpha + 1})$ for some $\alpha>0$.

\subsection{Asymptotic convergence in probability}
The consistency of the estimator is established in \cite{MMfull} under four sets of assumptions which are recalled here for completeness. 
Assumption~\ref{ass:identifiability} has been given in Section~\ref{sec:identif} to ensure the identifiability of the underlying model. Assumption~\ref{ass:regularite} state the constraints on the distribution of the components. Assumption~\ref{ass:penalty} gives some conditions on the penalty term. Finally, Assumption~\ref{ass:intervals} gives some conditions on the discretization. 

\begin{ass}\label{ass:regularite}
\begin{enumerate}[label=(\roman*)]
\item \label{ass:regularite1} There exists some function $\tau$ in $L_1(p_0 \nu)$ such that, for any model $\bm \in \mathcal{M}$ and any parameter $ \bth \in \Theta_{\bm}$, $$|\ln p(.|\bm, \bth) |<\tau \; \; \nu\text{-a.e.} .$$
\item \label{ass:regularite3} Each variable $j$ is defined on a compact space $\mathcal{X}_j$ and its densities for each component $g$, denoted by $\eta_{gj}$, are strictly positive except on a set of Lebesgue measure zero.
\item \label{ass:regularite2} There exists some positive constant $L<\infty$ which, for any block $b$ and any variable $j \in \Omega_b$, bounds the derivative of the densities $\eta_{gj}$ over $\mathcal{X}_j$:  $$\forall x_j\in\mathcal{X}_j, \; \; |\eta_{gj}'(x_j)|\leq L.$$ 
\end{enumerate}
\end{ass}

\begin{ass}\label{ass:penalty}
\begin{enumerate}[label=(\roman*)]
\item \label{ass:penalty1} For any model $\bm$, $a_{n,\bm,R}$ is an increasing function of $R$, $G_1, \ldots, G_B$, $|\Omega_1|, \ldots, \text{ and } |\Omega_B|$.
\item \label{ass:penalty2} For any model $\bm$, $a_{n,\bm,R}/n$ tends to 0 as $n$ tends to infinity.
\item \label{ass:penalty3} For any model $\bm$, $R/a_{n,\bm,R}$ tends to 0 as $n$ tends to infinity.
\item \label{ass:penalty4} For any models $\bm$ and $\widetilde{\bm}$ with $\bm \subset \widetilde{\bm}$, the limit inferior of $a_{n,\widetilde{\bm},R}/a_{n,\bm,R}$ is strictly larger than one as $n$ tends to infinity.
\end{enumerate} 
\end{ass}

\begin{ass}\label{ass:intervals}
\begin{enumerate}[label=(\roman*)]
\item The number of bins $R$ tends to infinity with $n$ in the following way $\lim_{n\to\infty} R=\infty$ and $\lim_{n\to\infty} R(\ln^2 n)/n^{1/2}=0$.\label{ass:vitesseB}
\item \label{ass:intervals2} 
The length of each bin is not zero and  satisfies, for any variable $j$ and any bin $r$, $|I_{Rjr}|^{-1}=O(R)$.
\item \label{ass:intervals1} For any variable $j$, let $\mathcal{I}_{jR}$ be the set of the upper bounds of the $R$ intervals, then, for any $x_j\in\mathcal{X}_j$,  $d(x_j,\mathcal{I}_{jR})$ tends to zero as $R$ tends to infinity. 
\end{enumerate}
\end{ass}

\begin{theorem} \label{thm:cvproba}
Assume that independent data arise from \eqref{eq:model} with true model $\bm_0=\{B_0, \bG_0, \bomega_0\}$, and that the set of competing models $\mathcal{M}$ is given by \eqref{eq:competing}. If, in addition,  Assumptions~\ref{ass:identifiability}, \ref{ass:regularite}, \ref{ass:penalty} and \ref{ass:intervals} hold true, then, the estimator $\widehat{\bm}_{n,R}$ defined by \eqref{eq:mhat} converges in probability to $\bm_0$, as $n$ goes to infinity.
\end{theorem}

The proof follows the same lines as Theorem 1 in \cite{MMfull}, by noticing that we deal with each block separately and we have to distinguish in the set of variables  which are involved in the considered block and which are not. 

\subsection{EM algorithm for model selection}
In order to compute $\widehat{\bm}_{n,R}$, we need to maximize the penalized log-likelihood over both $\bth_R$ and $\bm$. We will make use of the complete-data log-likelihood, which is based on the supposed observation of the component membership $\tz$ and is thus defined by 
\begin{equation}
\ell(\bth_R|\bm,\tx,\tz)= \sum_{b=1}^B \ln p(\tz_b|G_b, \bpi_{b}) + \sum_{b=1}^B \sum_{j \in \Omega_b} \ln p(\tx_{j}|G_b, \tz_{b}, \balpha_{Rbj}), 
\end{equation}
where $\tx_{j} = (x_{1j},\ldots,x_{nj})$, $\balpha_{Rbj}=\{\alpha_{Rgjr}, g=1, \ldots, G_b, r=1, \ldots, R\}$, $$\ln p(\tz_{b}|G_b, \bpi_{b})=\sum_{i=1}^n \sum_{g=1}^{G_b} z_{ibg} \ln \pi_{bg},$$ and
$$ \ln p(\tx_{j}|G_b, \tz_{b}, \balpha_{Rj}) = \sum_{i=1}^n \sum_{g=1}^{G_b} z_{ibg} \sum_{r=1}^R \sigma_{Rjr}(x_{ij}) \ln \left( \frac{\alpha_{Rgjr}}{|I_{Rjr}|} \right).$$
The maximum likelihood estimates (MLE) can be obtained by an EM algorithm \citep{Dem77,Mcl97}. Independence between the $B$ blocks of variables permits to maximize the observed-data log-likelihood on each block independently.

Due to the number of competing models, an exhaustive approach which consists in computing BIC for each competing models is not doable in practice. We will use the same idea as in \cite{MMVV} to circumvent the combinatorial issue. Holding $(B,\bG)$ fixed, model selection with BIC and maximum likelihood inference implies to maximize the penalized likelihood with respect to $(\bomega,\bth_R)$. 
This maximization can be carried out simultaneously by using a modified version of the EM algorithm \citep{green1990use, marbac2017variable2} and, then, $\widehat{\bm}_{n,R}$ can be found by running this algorithm for each value  of $(B,\bG)$ allowed by $\MM$. 
Therefore, less than $\sum_{B=1}^{B_{\max}} G_{\max}^{B}$ calls of the EM algorithm should be done.  Note that the number of calls of EM algorithm does not depend on the number of variables and this not intensive if one considers $B_{\max}$ small (\emph{i.e.,} $B_{\max}<5$). This can seem restrictive, but note that classical clustering methods consider $B_{\max}=1$. 
Moreover, if $B_{\max}$ is wanted to be more than five, then the model remains well defined but the proposed method for model selection suffers from combinatorial issues. Then, in this case, other algorithms (like forward/backward search) should be used for model estimation. 

To implement this modified EM algorithm, we introduce the penalized complete-data log-likelihood
\begin{equation*}
\ell_{pen}(\bth_R|\bm,\tx,\tz)= \ell(\bth_R|\bm,\tx,\tz) - a_{n,\bm,R}.
\end{equation*}
We need to assume that the penalty can be decomposed as a sum of penalties concerning each type of parameter as follows:
$a_{n,\bm,R}= \sum_{b=1}^B \left(a_{n,\bpi_b,R} + \sum_{j \in \Omega_b} a_{n,\balpha_{Rbj},R} \right).$ This assumption is not stringent and is satisfied, for instance, by the BIC with $ 2 a_{n,\bpi_{b},R}= (G_b-1) \ln n $ and $2 a_{n,\balpha_{Rbj},R} = (R-1)(G_b-1) \ln n $.
The penalized complete-data log-likelihood thus rewrites as
\begin{equation}\label{eq:pencompl}
\ell_{pen}(\bth_R|\bm,\tx,\tz) = \sum_{b=1}^B \left( \ln p(\tz_{b}| \bpi_{b}) - a_{n,\bpi_b,R} \right)   + \sum_{b=1}^B \sum_{j\in\Omega_b}\left( \ln p(\tx_{j}| \tz_{b}, \balpha_{Rbj}) - a_{n,\balpha_{Rbj},R} \right).
\end{equation}

Holding $(B,\bG)$ fixed and starting from $(\bomega^{[0]},\bth_R^{[0]})$, the iteration $[s]$ of the algorithm is composed of three steps:\\
\textbf{E-step} Computation of the fuzzy partitions $t_{ibg}^{[s]}:=\mathbb{E}[Z_{ibg}|\bx_i,\bm^{[s-1]},\bth_R^{[s-1]}]$, hence for $b=1,\ldots,B$, for $g=1,\ldots,G_b$, for $i=1,\ldots,n$
$$t_{ibg}^{[s]}=\dfrac{\pi_{bg}^{[s-1]} \prod_{j \in \Omega_b^{[s-1]} } \left(\alpha_{Rgjr}^{[s-1]}\right)^{\sigma_{Rjr}(x_{ij})}}{\sum_{k = 1}^{G_b} \pi_{bk}^{[s-1]} \prod_{j \in \Omega_b^{[s-1]}} \left(\alpha_{Rkjr}^{[s-1]}\right)^{\sigma_{Rjr}(x_{ij})}}.$$
\textbf{M-step1} Updating the affectation of the variables to blocks, for $j=1,\ldots,d$,
$$ \omega_{j}^{[s]}= \argmax_{b \in\{1,\ldots,B\}} \left(\sum_{g=1}^{G_{b}}\max_{\balpha_{Rgj} \in 
S_{R}}  Q(\balpha_{Rgj}|\tx_{j},\ttt_{bg}^{[s]}) - a_{n,\balpha_{Rbj},R}  \right),$$
where $\balpha_{Rgj}=(\alpha_{Rgj1}, \ldots, \alpha_{RgjR})$ and $Q(\balpha_{Rgj}|\tx_{j},\ttt)= \sum_{i=1}^n t_i \sum_{r=1}^R \sigma_{Rjr}(x_{ij})  \ln \alpha_{Rgjr}$.
Thus $\Omega_b^{[s]}= \{j : \omega_{j}^{[s]}= b \}$.\\
\textbf{M-step2} Updating the model parameters: for $b=1,\ldots,B$, for $g=1,\ldots,G_b$,
$$\pi_{bg}^{[s]}=\frac{1}{n} \sum_{i=1}^n t_{ibg}^{[s]}$$
 and for $j \in \Omega_b^{[s]}$,  
$$\balpha_{Rgj}^{[s]} = \argmax_{\balpha_{Rgj} \in S_{R}} Q(\balpha_{jg}|\tx_{j},\ttt_{\omega_j^{[s]}g}^{[s]}),$$
which implies that for any $r=1, \ldots, R$,
$$\alpha_{Rgjr}^{[s]}= \frac{\sum_{i=1}^n t_{i \omega_j^{[s]}g} \sigma_{Rjr}(x_{ij})}{\sum_{i=1}^n t_{i \omega_j^{[s]}g}}.$$

Like for the standard EM algorithm, the objective function $\ell_{pen}(\bth_R|\bm,\tx,\tz)$ increases at each iteration but the global optimum is not achieved in general. Hence, different random initializations must be done. 
Finally, note that the algorithm can return empty blocks. 
Indeed, M-step1 is done without constraining each block to contain at least one variable. 
Thus, each $ \omega_{j}^{[s]}$ can be obtained independently. 

This algorithm permits to estimate the densities of the underlying model. However, the bin-based density estimators are generally proven to be outperformed by kernel-based density estimators. Thus, we advise to use the proposed method only for selecting the model, and then to use kernel-based density estimates for the selected model, for instance EM-like algorithm \citep{BenagliaJCGS2009} or MM-algorithm \citep{LevineBiometrika2011}. This will be illustrated in the numerical experiments.

\section{Numerical experiments}\label{sec:experiments}

\subsection{Simulation setup}

Data are generated from a mixture with three blocks of variables ($B=3$), each block being a mixture of three components ($\bG = (3,3,3)$ )  with well-balanced clusters (\emph{i.e.,} with equal proportions $\pi_{bg}=1/3$) and with the same number of involved variables (\emph{i.e.,} $|\Omega_1|= |\Omega_2| = |\Omega_3|$). In block $b$, the marginal density of $X_i$ given the component membership is a product of $|\Omega_b|$ univariate densities such that $$X_{ij}=  \sum_{g=1}^{G_b} z_{ibg} \delta_{gj} + \xi_{ij}$$ where all the $\xi_{ij}$ are independent and where  $\delta_{gj}=\tau$ if $j$ is equal to $g$ modulo $G_b$ and $\delta_{gj}=0$ otherwise. The value of $\tau$ is tuned in order to obtain a chosen theoretical misclassification rate. 

We consider three distributions for the $\xi_{ij}$ (standard Gaussian, Student with three degrees of freedom and Laplace), four sample sizes ($n=50, 100, 200, 400$), different numbers of variables in each block ( for each block $b$, $|\Omega_b|$ equals successively 6, 9 and 12) while the value of $\tau$ is defined to obtain a theoretical misclassification rates of $5\%$ and of $10\%$.
 For each situation ( sample size, number of variables, law and misclassification rate), 100 replicates are generated. 

The discretization step is conducted with $R$ bins given by the empirical quantiles. We investigate four number of bins: $R= [n^{1/4}], [n^{1/5}], [n^{1/6}] $ and $[n^{1/7}]$, for each previous replicate.

The model selection is done with the proposed modified EM algorithm, with $B_{\text{max}} = 3$ and $G_{\text{max}} = 3$, and with a BIC penalty. Then, for the selected model, parameters and density estimation is conducted with the package \texttt{mixtools}, leading to a more accurate estimated partition.

\subsection{Performances of the proposed method.}
We compute in each situation the Adjusted Rand Index (ARI) between the obtained blocks of variables and the true ones, as well as the ARI between the obtained partition of the individuals and the true partition, in each block.

\begin{figure}[htp!]
    \centering
    \includegraphics[scale=0.4]{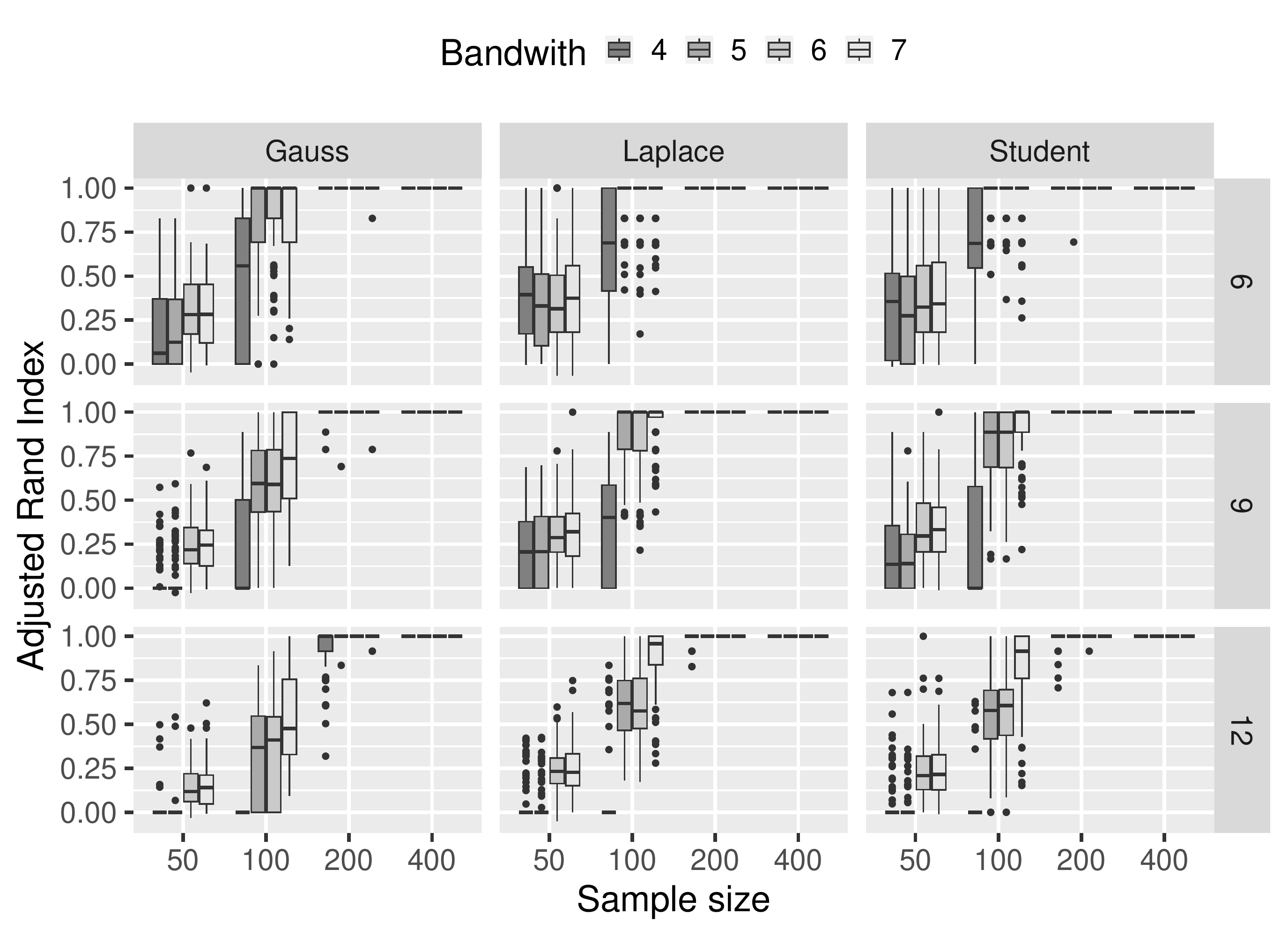}
    \caption{ARI between the true and the estimated blocks of variables, for different sample size, different families of components, different number of variable, and for different number of bins  $[n^{1/7}]$, $[n^{1/6}]$, $[n^{1/5}]$, $[n^{1/4}]$ indicated by grey levels.}
    \label{fig:ARI_om_compB}
\end{figure}

Figure \ref{fig:ARI_om_compB} displays the boxplot of the ARI between the true and estimated blocks of variables obtained on each situation over the 100 replicates. The misclassification rate is fixed equal to $5\%$. It shows that the proposed method is able to perfectly recover the different blocks of variables, for each family of law, as soon as the sample size and the number of bins are sufficiently large. We recall that the number of bins should not be too large compared to the sample size, in order to satisfy Assumption \ref{ass:intervals}. The method is more accurate as $n$ grows, while it deteriorates a little when the number of variables in each blocks grows, for small values of $n$. This is not surprising as it makes the classification problem more complex. This will be further illustrated on the last figures.

\begin{figure}[htp!]
    \centering
    \includegraphics[scale=0.4]{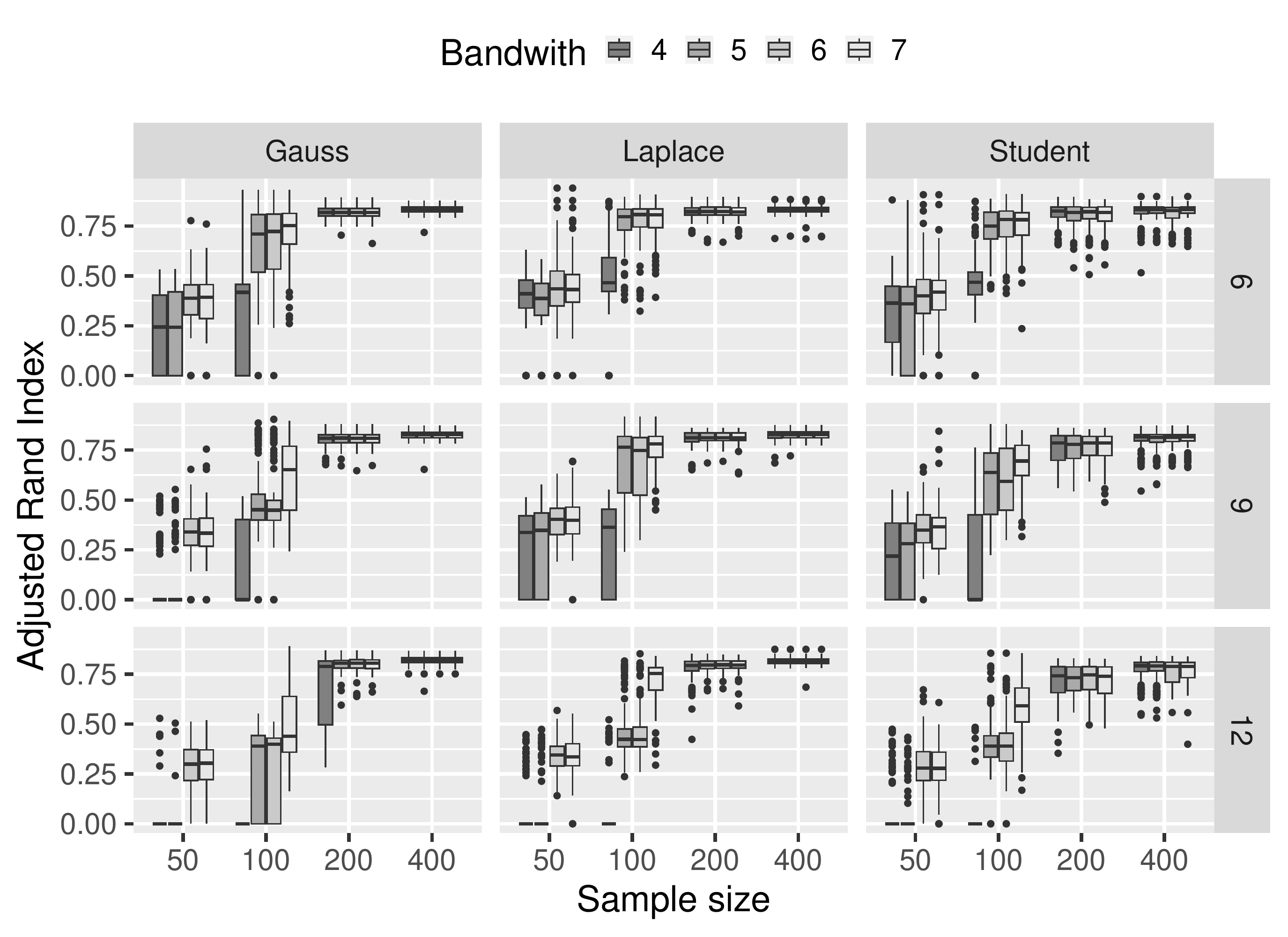}
    \caption{mean ARI between the true and the estimated partition of individuals, for different sample size, different families of components, different number of variable, and for different number of bins  $[n^{1/7}]$, $[n^{1/6}]$, $[n^{1/5}]$, $[n^{1/4}]$ indicated by grey levels.}
    \label{fig:ARI_compB}
\end{figure}

Figure \ref{fig:ARI_compB} displays the boxplots over of the mean of the ARI obtained in each block between the true and estimated partitions. The misclassification rate is fixed equal to $5\%$. Here, one can notice that the performance are good as well. 

\begin{figure}[htp!]
    \centering
    \includegraphics[scale=0.4]{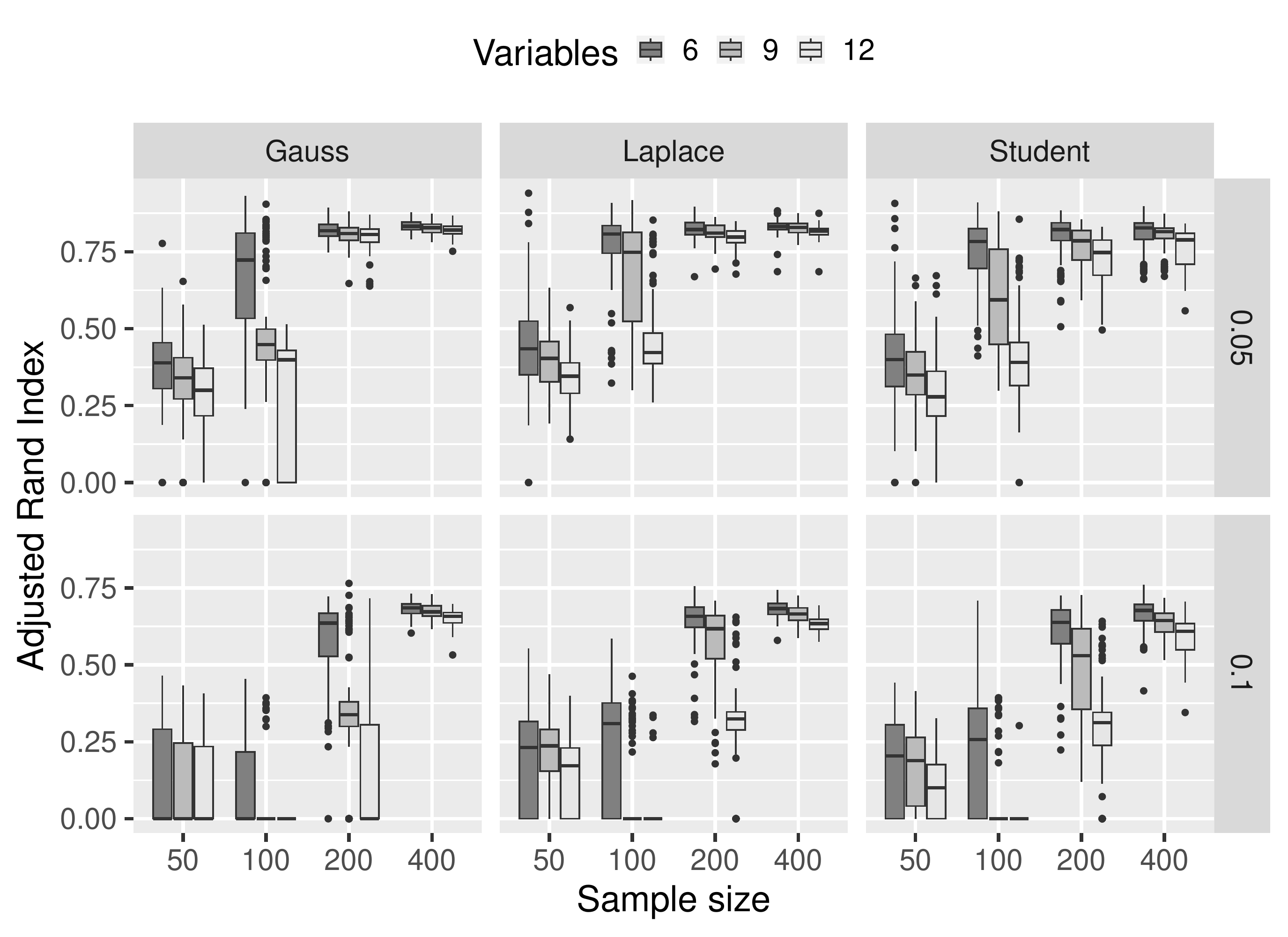}
    \caption{mean ARI between the true and the estimated partition of individuals, for different sample size, different families of components, different misclassification rate and different number of variables indicated by grey levels.}
    \label{fig:ARI_moy}
\end{figure}

\begin{figure}[htp!]
    \centering
    \includegraphics[scale=0.4]{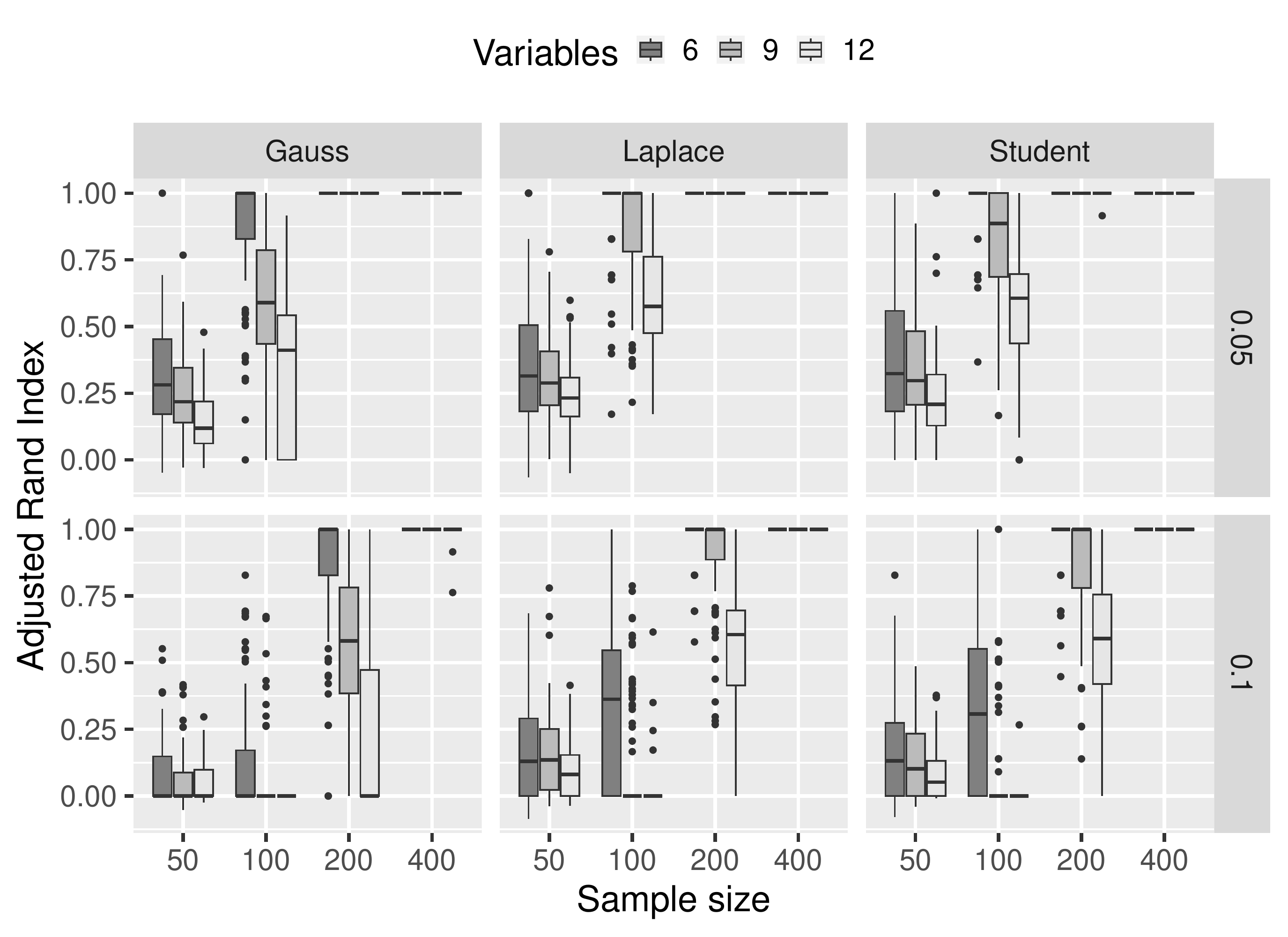}
    \caption{mean ARI between the true and the estimated blocks of variables, for different sample size, different families of components, different misclassification rate and different number of variables indicated by grey levels.}
    \label{fig:ARI_om}
\end{figure}

In Figures \ref{fig:ARI_moy} and \ref{fig:ARI_om}, we investigate the behavior of the proposed method, for different values of the theoretical misclassification rate. The number of bins is fixed equal to $[n^{1/6}]$. Again, we note that the methods is more accurate as $n$ grows, while it deteriorates with the increase of the number of variables in each block. 

\section{Conclusion}
We have proposed a new method for performing clustering with multiple partitions, when no parametric assumptions are made on the conditional distributions of the variables given the component memberships. This methods permits do deal with continuous data. It is based on the discretization of the continuous data which permits to reuse previous methods which are able to deal with multinomials distributions. For mixed-type data, the method can be straightforwardly extended by discretizing only the continuous ones. Model selection is conducted on the discretized data by using a modified EM algorithm, which estimates simultaneously the partition of the variables and the parameters of the model. The procedure is consistent for a range of penalty including the classical BIC.

\section*{Bibliography}

\bibliography{biblio}
\bibliographystyle{apalike}

\appendix

\end{document}